\documentclass[11pt]{article}
\usepackage[top=1.in,bottom=1.in,left=1.in,right=1.in]{geometry}
\usepackage{graphicx}
\usepackage{wrapfig} 
\usepackage{subfig} 
\usepackage{amsmath}
\usepackage{color}
\usepackage[pagewise]{lineno}
\graphicspath{{figures/}}
\DeclareGraphicsExtensions{.pdf,.png,.jpg}

\newcommand{\nub}{\bar{\nu}}
\newcommand{\nue}{\nu_e}
\newcommand{\nueb}{\bar{\nu}_e}
\newcommand{\numu}{\nu_\mu}
\newcommand{\numub}{\bar{\nu}_\mu}

\begin{document}
\title{Letter of Intent: 
A new investigation of $\numu\rightarrow\nue$ oscillations 
with improved sensitivity in an enhanced MiniBooNE experiment}
\author{
\large{\textbf{MiniBooNE Collaboration}} 
\vspace*{5pt} \\
R. Dharmapalan, S. Habib, C. Jiang, \& I. Stancu \\
\textit{University of Alabama, Tuscaloosa, AL 35487} \vspace{3pt} \\
Z. Djurcic \\
\textit{Argonne National Laboratory, Argonne, IL 60439} \vspace{3pt} \\
R. A. Johnson \& A. Wickremasinghe \\
\textit{University of Cincinnati, Cincinnati, OH 45221} \vspace{3pt} \\
G. Karagiorgi \& M. H. Shaevitz \\
\textit{Columbia University, New York, NY 10027} \vspace{3pt} \\
B. C. Brown, F.G. Garcia, R. Ford, W. Marsh, C. D. Moore, \\
D. Perevalov,  \& C. C. Polly \\
\textit{Fermi National Accelerator Laboratory, Batavia, IL 60510} \vspace{3pt} \\
J. Grange,  J. Mousseau, B. Osmanov, \& H. Ray \\
\textit{University of Florida, Gainesville, FL 32611} \vspace{3pt}\\
R. Cooper, R. Tayloe \\
\textit{Indiana University, Bloomington, IN 47405}\vspace{3pt}  \\
G. T. Garvey, W. Huelsnitz, W. C. Louis, C. Mauger, G. B. Mills, \\
Z. Pavlovic, R. Van de Water, \& D. H. White \\
\textit{Los Alamos National Laboratory, Los Alamos, NM 87545} \vspace{3pt} \\
R. Imlay, M. Tzanov \\
\textit{Louisiana State University, Baton Rouge, LA 70803}\vspace{3pt} \\
L. Bugel, J. M. Conrad \\
\textit{Massachusetts Institute of Technology; Cambridge, MA 02139} \vspace{3pt} \\
B. P. Roe \\
\textit{University of Michigan, Ann Arbor, MI 48109} \vspace{3pt} \\
A. A. Aguilar-Arevalo \\
\textit{Instituto de Ciencias Nucleares, Universidad Nacional Aut\'onoma de M\'exico,} \\
\textit{M\'exico D.F. M\'exico} \vspace{3pt} \\
P. Nienaber \\
\textit{Saint Mary's University of Minnesota, Winona, MN 55987}\\
}

\maketitle
\thispagestyle{empty}

\clearpage
\newpage

\begin{abstract}
We propose adding 300~mg/l PPO to the existing MiniBooNE detector mineral 
oil to increase the scintillation response.  This will allow the detection  
of associated neutrons and increase sensitivity to final-state nucleons
in neutrino interactions. This increased capability will enable an independent
test of whether the current excess seen in the MiniBooNE oscillation search is
signal or background.  In addition it will enable other neutrino interaction
measurements to be made including a search for the strange-quark contribution 
to the nucleon
spin ($\Delta s$) and a low-energy measurement of charged-current quasielastic
scattering. 
\end{abstract}

\tableofcontents
\clearpage
\newpage

\section{Executive Summary}
In this Letter, we describe a plan for adding scintillator to 
the existing MiniBooNE mineral oil detector medium to allow  
a test of the neutral-current/charged-current (NC/CC) nature of the MiniBooNE
low-energy excess.  The scintillator will enable reconstruction of 
2.2~MeV $\gamma$ from neutron-capture on protons following neutrino 
interactions.   Charged-current neutrino interactions,
at low-energy where the MiniBooNE oscillation excess is observed, should
have associated neutrons with less than 10\% probabability. This is in 
contrast to the neutral-current backgrounds which should have associated
neutrons in approximately 50\% of events.  These predicted neutron 
fractions will be measured in MiniBooNE for both the charged- and 
neutral-current channels, thereby eliminating that systematic uncertainty.
A measurement of the neutron-fraction in a new appearance oscillation 
search with MiniBooNE will increase the significance of the oscillation
excess, if it maintains in the new data set, to at minimum $5\sigma$.   

This new phase of MiniBooNE will enable additional important 
studies such as the spin structure of nucleon ($\Delta s$) via NC elastic
scattering, a low-energy measurement of the neutrino flux via 
$\numu~^{12}C \rightarrow \mu^{-}~^{12}N_\textrm{g.s.}$ scattering, and a test of the
quasielastic assumption in neutrino energy reconstruction.  These topics
will yield important, highly-cited results over the next 5 years for 
a modest cost, and will help to train Ph.D. students and postdocs.  

This enterprise will offer complementary information to that from 
the upcoming MicroBooNE experiment. MicroBooNE uses an Argon target,
and it will be helpful to further study the MiniBooNE excess with
new methods but using the same carbon nucleus. In addition, 
the beam will be delivering neutrinos for MicroBooNE, so there is
minimal incremental cost to also deliver beam to MiniBooNE.

\begin{quotation}
\textit{This program of measurements requires approximately $6.5\times10^{20}$ 
protons on target delivered to MiniBooNE  and can begin in early 2014. 
We are requesting support of this concept to enable the collaboration 
to plan the experiment and analysis in more detail with the goal of submitting a full 
proposal for the experiment in mid-2013.}
\end{quotation}

\clearpage
\section{Introduction}
The MiniBooNE experiment has, for the last 10 years, searched
for $\numu\rightarrow\nue$ and $\numub\rightarrow\nueb$ 
in the Booster Neutrino Beamline at Fermilab.  The beam was shut down in April 2012
to enable the Fermilab accelerator complex to be upgraded in preparation for 
delivering higher beam power to the NOvA experiment.  Before the shutdown, 
MiniBooNE completed an antineutrino phase of running which brought the total 
amount of beam delivered to 
the experiment to $11.3\times 10^{20}$ protons on target (POT) in antineutrino
mode and $6.5\times 10^{20}$~POT in neutrino mode.  

\begin{figure}[h]
\centering
\begin{minipage}[t]{0.48\textwidth}
\centering
{\includegraphics[width=0.95\textwidth]{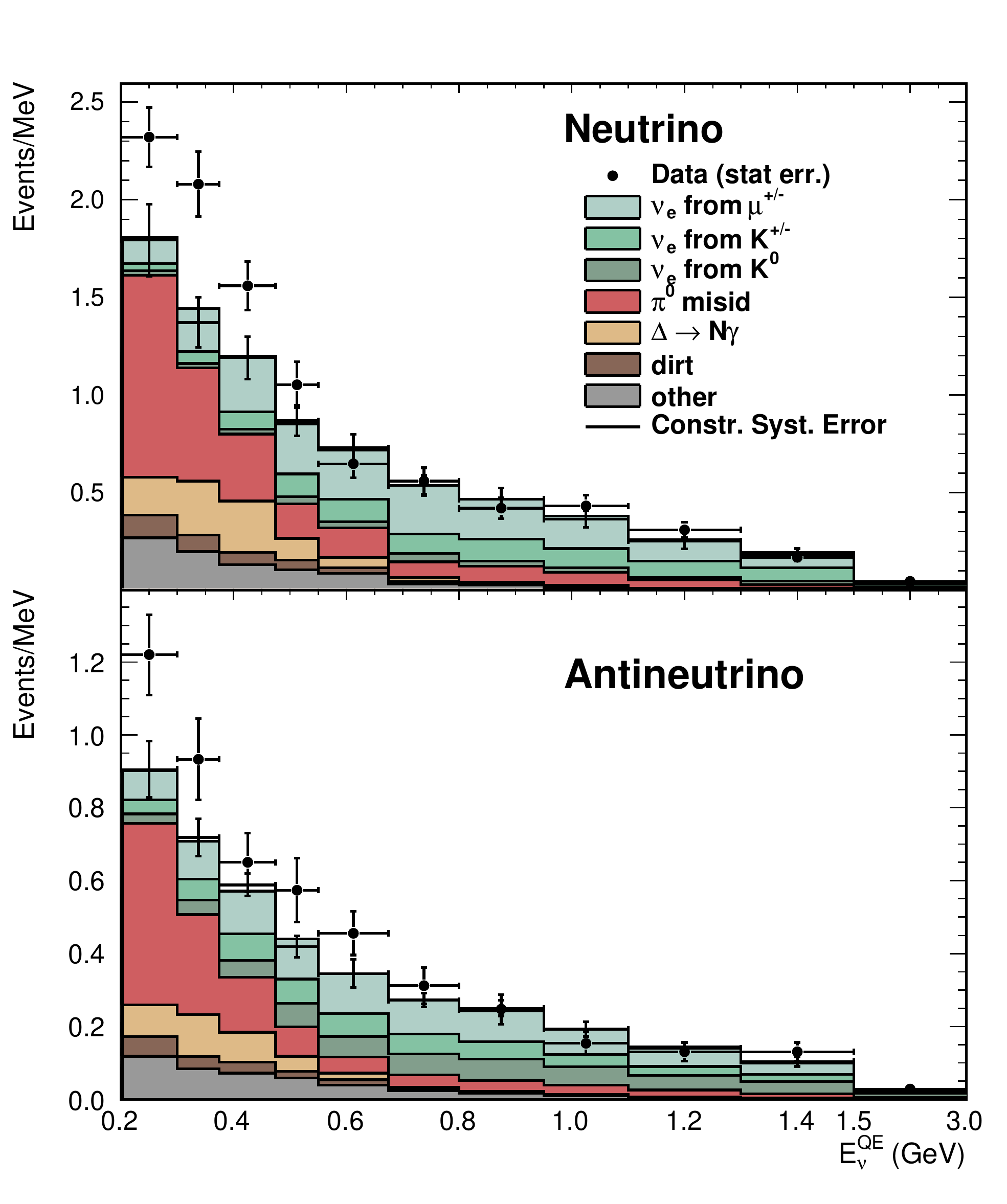}}
\caption{The neutrino mode (top) and antineutrino mode (bottom) 
reconstructed neutrino energy, $E_\nu^{QE}$, distributions   
for data (points with statistical errors) 
and predicted background (histogram with systematic errors).}
\label{fig:nu_nubar_stack}
\end{minipage}
\hspace{3mm}
\begin{minipage}[t]{0.47\textwidth}
\centering
{\includegraphics[width=0.95\textwidth]{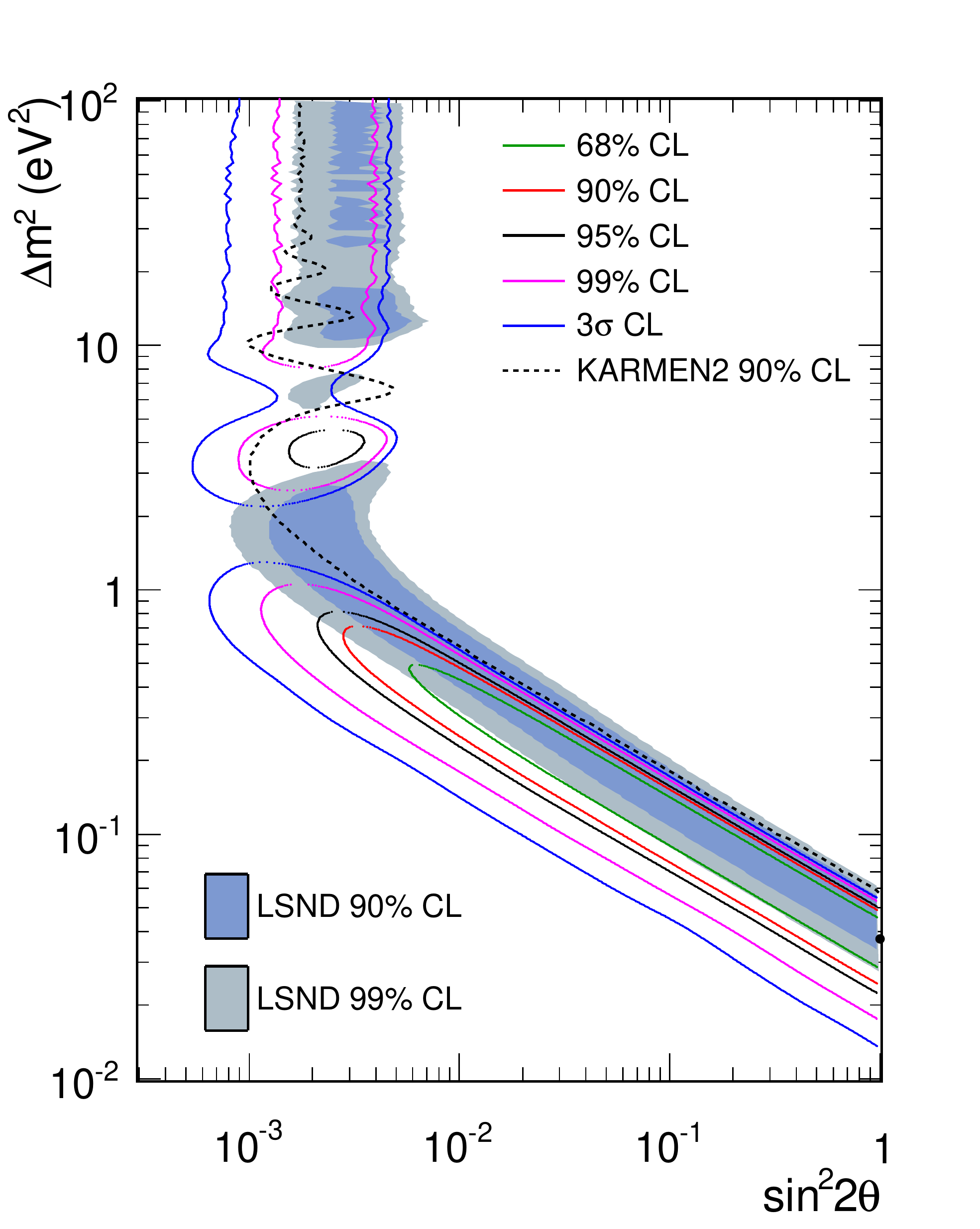}}
\caption{
MiniBooNE allowed regions in combined neutrino and antineutrino mode for events with
$200<E^{QE}_{\nu}< 3000$~MeV within a 2$\nu$ oscillation model. }
\label{fig:mb_nu_nubar_combined_fit}
\end{minipage}
\end{figure}

Both $\numu\rightarrow\nue$ and $\numub\rightarrow\nueb$ oscillation 
analyses have been conducted with this data 
individually~\cite{AguilarArevalo:2007it}-\cite{AguilarArevalo:2010wv}
and recently as a combined data set with the latest updates to the
antineutrino data~\cite{AguilarArevalo:2012va}.
There is an excess of events over the calculated background in both modes 
(Fig.~\ref{fig:nu_nubar_stack}) examined individually as well as for the combined 
data set which contains a total excess of  $240.3\pm62.9$ ($3.8\sigma$) events.  
A two-neutrino fit to the combined data set  yields allowed parameter 
regions (Fig.~\ref{fig:mb_nu_nubar_combined_fit}) which
are consistent with oscillations in the 0.01 to 1 eV$^2$ $\Delta m^2$ range 
and are consistent with the regions reported by the LSND experiment~\cite{Aguilar:2001ty}.

The excess occurs in both $\nu$ and $\nub$ samples at low energy, and so it is 
natural to consider more carefully the largest backgrounds in 
that energy region.  They are dominated by neutral-current (NC) $\pi^0$ and NC
$\Delta$ radiative decays ($\Delta \rightarrow N \gamma$). Both of these NC 
processes are constrained  by measurements within MiniBooNE, but an anomalous 
process such as NC $\gamma$  production, not sufficiently accounted for in the MiniBooNE 
analysis, could lessen the significance of the oscillation excess.  
We are proposing to measure these backgrounds with a new technique combined with
additional running of MiniBooNE. 

The MiniBooNE detector uses 800 tons of mineral oil
(CH$_2$) as a target medium for inducing CC and NC neutrino interactions.  
The mineral oil also serves as the detector medium for observing the final state 
particles resulting from the interactions.
This is achieved via detection of the Cerenkov light from charged particles in the 1280 
8'' photomultiplier tubes (PMTs) that line the inside of the spherical
detector tank.   In addition to the Cerenkov light produced in a cone
around the trajectory of charged particles, some isotropic 
scintillation light is produced due to presence of aromatic impurities
in the mineral oil.  

We propose adding approximately 300~kg of PPO scintillator to the 800 tons 
of MiniBooNE mineral oil to increase the amount of scintillation light produced
by 2.2~MeV $\gamma$ that result from delayed ($\tau\approx186~\mu$s) neutron 
capture on protons within the mineral oil.  This will allow an important test of the 
oscillation signal by checking that the excess is indeed due to CC interactions
of low-energy neutrinos and not an incorrectly calculated NC background.  This 
can be done by counting  $n$-capture events that follow oscillation candidate
events.  If the excess is indeed due to CC interactions of low energy $\nue$, 
only approximately $10\%$ of the excess will have associated $n$-capture events.  
If, instead, the excess is due to a NC process, one would expect many
more neutrons produced since the interactions are caused from
higher energy neutrinos.  One expects approximately 50\% of NC
background events to have an associated neutron.  An attractive
feature of this method is that the neutron fraction for CC and NC
processes may be measured with MiniBooNE via similar channels thereby
eliminating that systematic uncertainty.

The increased level of scintillation will enable several other important
measurements.  The detection of $n$-capture enables a measurement of
the neutron to proton ratio in NC elastic scattering which is sensitive to
the strange-quark spin of the nucleon ($\Delta s$).  The  $\beta$ decay from 
the $^{12}N_\textrm{g.s.}$ in the $\numu~^{12}C \rightarrow \mu^{-}~^{12}N_\textrm{g.s.}$ process
will be better reconstructed which will allow a measurement of this process
and a check of the low-energy neutrino flux.   Low-energy recoil nucleons
will be more visible within neutrino events allowing a test of the quasielastic
assumption in neutrino energy reconstruction.

In the following sections, we describe the extended physics program that is made possible
with the addition of scintillator to MiniBooNE. We will discuss how the 
scintillation light will be increased,  how it affects event reconstruction, 
and describe a plan for future running with MiniBooNE 
in this new configuration.

\clearpage
\section{Physics Goals}
A main motivation for adding scintillator to MiniBooNE is to provide a test of 
the nature of the low-energy excess of events observed in both the $\nue$ and $\nueb$ appearance 
searches conducted by MiniBooNE.  The addition of scintillator will also enable 
an investigation of the strange-quark contribution  to the nucleon spin ($\Delta s$), 
a measurement of the $\numu~^{12}C \rightarrow \mu^{-}~^{12}N$ reaction, and a test 
of the quasielastic assumption in neutrino energy reconstruction. 

\subsection{Oscillation search with CC/NC identification}

MiniBooNE has measured a $3.8\sigma$ excess of oscillation candidate events in the combined
$\numu$ and $\numub$ data sets collected to date at Fermilab~\cite{AguilarArevalo:2012va}.
As can be seen in Fig.~\ref{fig:nu_nubar_stack}, the predicted backgrounds in the 
low energy regions, where the excess is most substantial, are dominated by neutral current
backgrounds.  These are from two major sources, both from NC interactions:  misidentification 
of the $\pi^0$ (``$\pi^0$ misid'') and the production of $\Delta$ baryons which then radiatively
decay (``$\Delta \rightarrow N\gamma$'').   A test of these NC backgrounds in a measurement 
with different systematic errors would be quite valuable to firmly establish the oscillation excess.

\begin{figure}[h]
\centering
\subfloat[$\nue CCQE$]{
\label{fig:fd_CCNC_set:a}
{\includegraphics[width=0.31\textwidth]{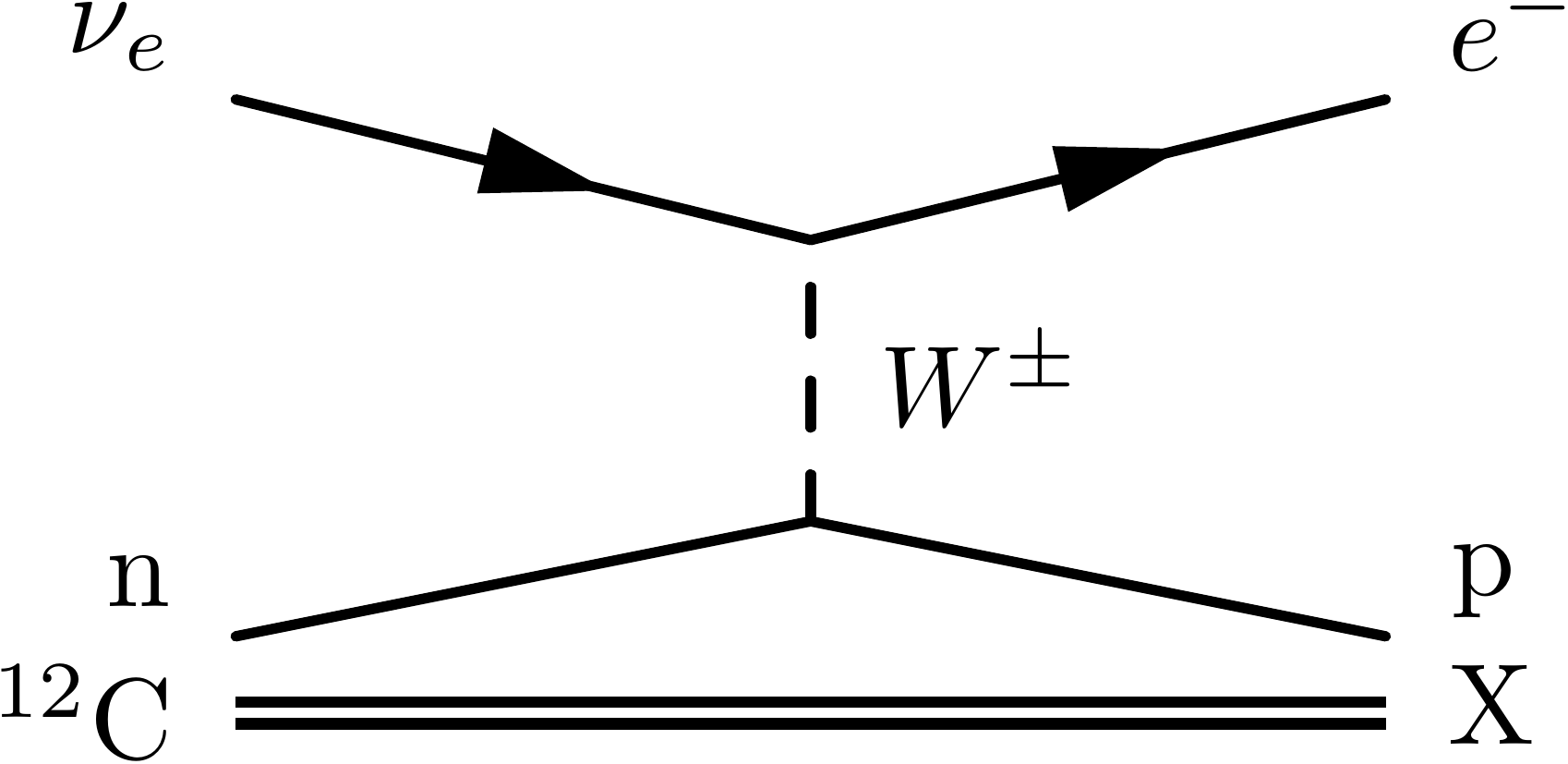}}
}
\hspace{-5mm}
\subfloat[$\numu NC \pi^0$ ]{
\label{fig:fd_CCNC_set:b}
{\includegraphics[width=0.31\textwidth]{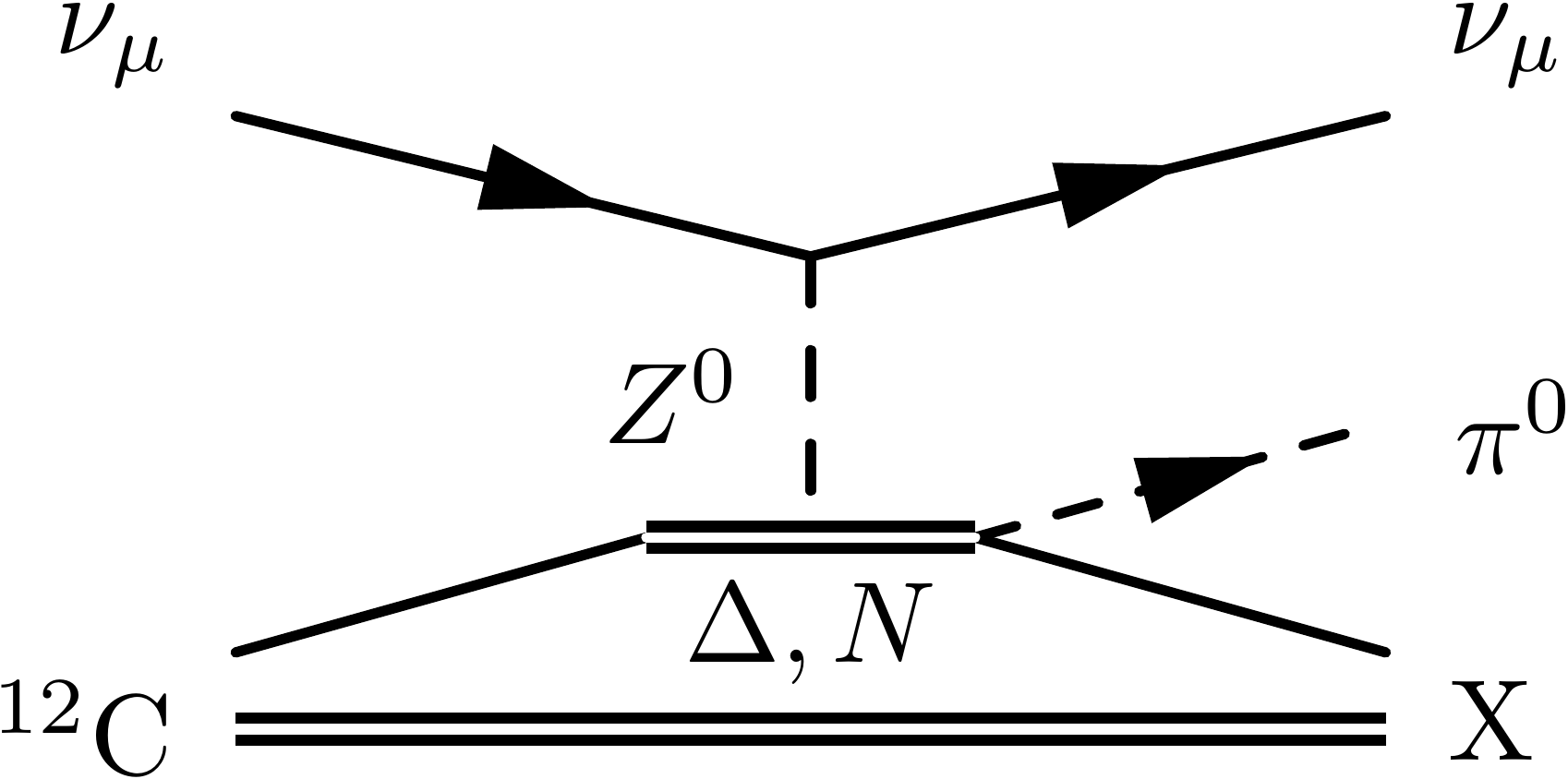}}
}
\hspace{-5mm}
\subfloat[$\numu NC \gamma$]{
\label{fig:fd_CCNC_set:c}
{\includegraphics[width=0.31\textwidth]{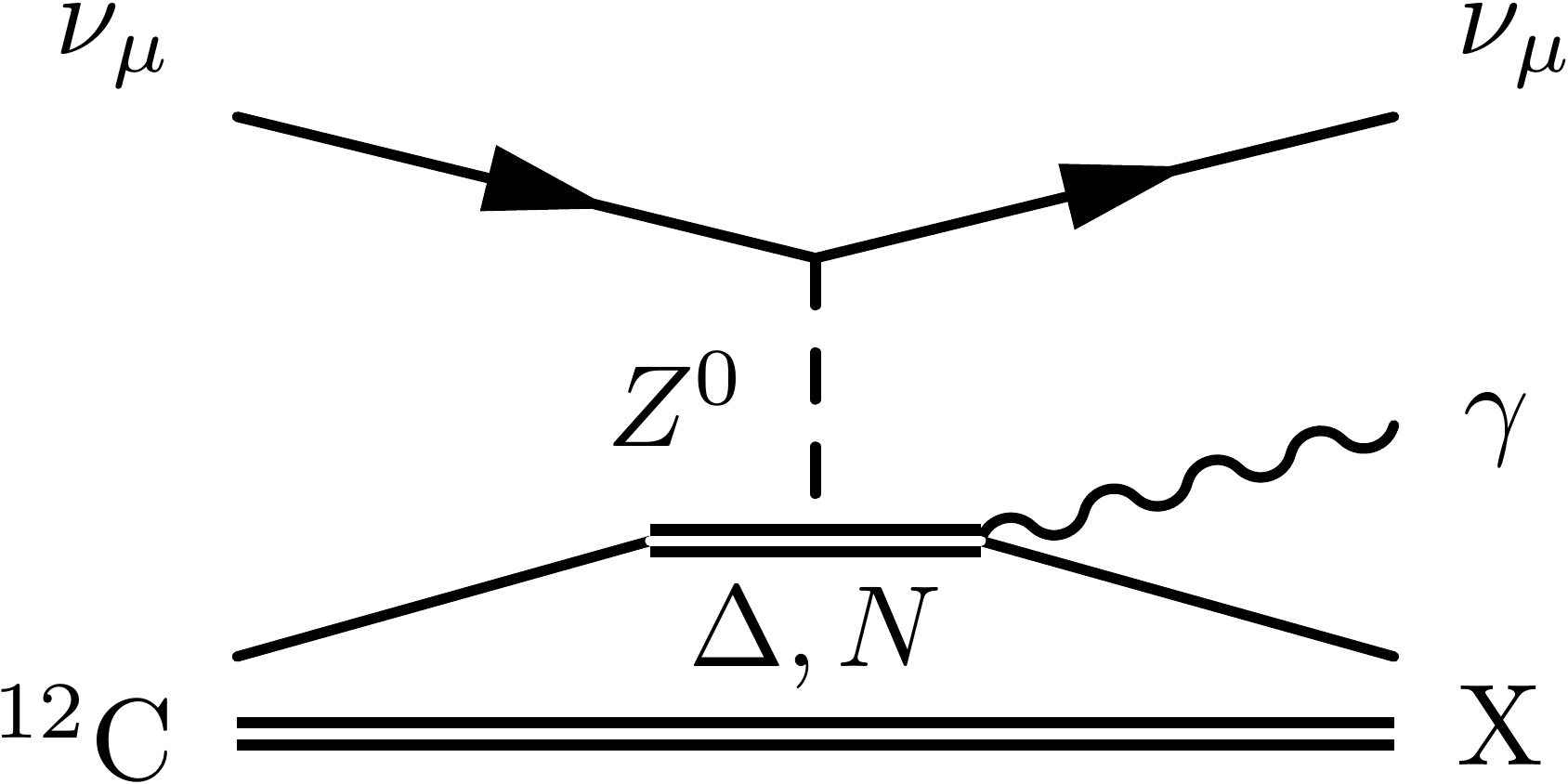}}
}
\caption{Diagrams of signal (a) and background (b,c) neutrino oscillation candidate events.}
\label{fig:fd_CCNC_set}
\end{figure}

MiniBooNE can perform this test by detecting neutrons associated with oscillation candidate
events. In brief, at low $E_\nu^{QE}$, true CC oscillation events (Fig.~\ref{fig:fd_CCNC_set:a}) 
should contain neutrons in less than 10\% of the events while the 
NC backgrounds (Figs.~\ref{fig:fd_CCNC_set:b},\ref{fig:fd_CCNC_set:c}) 
should contain neutrons  in $\approx 50\%$ of the events.

Note that  $E_\nu^{QE}$ is the reconstructed neutrino energy using the assumption
of neutrino quasielastic scattering from a neutron. This quantity should be a good 
estimate for the actual neutrino energy in true CC oscillation events (excepting possible
nuclear effects, Sec.~\ref{sec:enuqetest}).
However, because of the large missing energy in NC events, the actual neutrino energy is,
on average, much higher than $E_\nu^{QE}$ for NC backgrounds events.  So, in a first approximation, the NC 
backgrounds will contain more final state neutrons because the events are from higher true neutrino
energy. More energy is transferred to the nucleus which causes more neutron
production as compared to the CC signal in which the true neutrino energy is lower and 
(neglecting final state interactions) produces a single proton. 

In practice, we would rerun the MiniBooNE oscillation search in neutrino mode after
the addition of scintillator in order to enable neutron detection. Oscillation candidates
would be selected with the same strategy as the original search.  From this sample, we would 
search for neutron capture events and measure the neutron fraction which would test the
NC background estimates. An important feature of this measurement is that the neutron fraction
may be ``calibrated'' for the oscillation search via separate MiniBooNE $\numu$ CCQE and 
$\numu$ NC$\pi^0$ measurements which
greatly reduces errors from any nuclear model uncertainties.
A quantitative study that includes a more complete nuclear physics model,
estimates of final-state interactions, detector efficiency, and an estimate of sensitivity 
is presented in Section~\ref{sec:analysis}.

\subsection{Proton to neutron ratio in NC elastic events}
The NC neutrino-nucleon elastic scattering (NC elastic) interaction, $\nu N \rightarrow \nu N$, 
is sensitive to the isoscalar-axial structure of the nucleon~\cite{Garvey:1993sg}, so  
will be sensitive to the effects of strange-quark contributions to the nucleon spin
($\Delta s$).  Therefore, the right type of measurement of NC elastic scattering
would contribute substantially to the nucleon spin puzzle, an area of continued interest and
effort (e.g.~\cite{Bass:2009dr}). This measurement of NC elastic scattering has not yet been realized. 

MiniBooNE has made the most accurate measurement to date of the differential cross section for
$\nu N \rightarrow \nu N$~\cite{AguilarArevalo:2010cx} and the analysis for the 
$\bar{\nu} N \rightarrow \bar{\nu} N$ process is almost complete~\cite{Dharmapalan:2011sa}. 
While these are valuable measurements to help with understanding of neutrino-nucleon scattering, 
they are not
sensitive to $\Delta s$ because the acceptance of MiniBooNE is approximately equal for 
neutrons and protons.   The $\nu p \rightarrow \nu p$ process is sensitive to $\Delta s$
with the opposite sign as $\nu n \rightarrow \nu n$ and any strange quark effects 
cancel in the existing MiniBooNE measurement.

This situation changes abruptly with the addition of neutron-capture tagging.  In that case,
the neutrons and protons can be separately identified and the neutron/proton
ratio,
\begin{equation}
R(p/n) = \frac{\sigma(\nu p \rightarrow \nu p)}{\sigma(\nu n \rightarrow \nu n)},
\label{eq:rpn}
\end{equation} 
is quite sensitive to $\Delta s$~\cite{Garvey:1993sg}.  Based on previous studies~\cite{Bugel:2004yk},
a rough estimate is that a 10\% measurement of $R(p/n)$ should result in an error of $\approx 0.05$ uncertainty
on $\Delta s$.  It should be realized that the recent results from MiniBooNE on the unexpectedly large 
CCQE cross section~\cite{AguilarArevalo:2010zc} may call into question the theoretical uncertainty involved 
in extracting $\Delta s$ from $R(p/n)$. If there are multinucleon correlations contributing substantially
to NC elastic scattering, it may not be clear how that affects the extraction of $\Delta s$. 
Regardless, a 10\% 
measurement of $R(p/n)$ will be a valuable constraint and will spur more theoretical investigation.

\subsection{A measurement of $\numu~^{12}C \rightarrow \mu^{-}~^{12}N_\textrm{g.s.}$}
The reaction $\numu~^{12}C \rightarrow \mu^{-}~^{12}N_\textrm{g.s.}$ is an interesting reaction
to study with a scintillator-enhanced MiniBooNE for several reasons.   
It comes with a  distinctive tag from the $\beta$-decay of the $^{12}N_\textrm{g.s.}$ with endpoint 
energy of 16.3~MeV and lifetime of 15.9~ms.  This addition of scintillator to MiniBooNE will allow
for high efficiency and better reconstruction of the $\beta$-decay.
Since it is an exclusive reaction, the theoretical cross section can be calculated to 
$\approx 2\%$ very near threshold~\cite{Athanassopoulos:1997rn}. It was measured by 
LSND for both $\numu$ and $\nue$~\cite{Athanassopoulos:1997rn,Athanassopoulos:1997rm} 
and by KARMEN for $\nue$~\cite{Bodmann:1992ur} to agree with theory to within experimental
errors.  A measurement by MiniBooNE of this theoretically 
well-known reaction would enable a test of the low-energy neutrino flux which could better constrain 
the low-energy oscillation excess.

The event signature is quite distinct.  The low-energy prompt $\mu^{-}$ and subsequent
decay $e^{-}$ would be detected with the usual techniques employed for $\numu$ CCQE events
combined with a requirement for a detected $\beta$-decay candidate.  With the 
addition of scintillator to make 2.2~MeV $\gamma$ visible, the efficiency for detecting
the 16.3~MeV-endpoint $\beta$ will be quite high.

The challenge is that the fraction of the total $\numu$ scattering events that interact
via $\numu~^{12}C \rightarrow \mu^{-}~^{12}N_\textrm{g.s.}$ is small.  In the lowest energy
bin at $E_\nu \approx 250$~MeV, the cross section is about 4\% of the $\numu$ CCQE cross section,
falling to about  0.5\% by $E_\nu \approx 400$~MeV~\cite{Kolbe:1999au}.  However, with the 
data sample proposed here the total $\numu$ event sample will be large, the $^{12}N_\textrm{g.s.}$ 
signature quite distinct, and an analysis will be worth the effort.

\subsection{A test of the QE assumption in neutrino energy reconstruction}
\label{sec:enuqetest}
MiniBooNE has reported absolutely normalized cross sections for various $\numu$-carbon processes
including  $\numu$ CCQE~\cite{AguilarArevalo:2010zc}, CC$\pi^+$~\cite{AguilarArevalo:2010bm}, 
CC$\pi^0$~\cite{AguilarArevalo:2010xt}, NC elastic~\cite{AguilarArevalo:2010cx}, 
and NC$\pi^0$~\cite{AguilarArevalo:2009ww}.  
They all show a 30-40\% larger cross section than predicted in previously existing 
models (e.g.~\cite{Caballero:2005sj}).   
One emerging idea is that two-nucleon correlations in carbon are contributing significantly to the
interaction cross section~\cite{Martini:2009uj,Martini:2010ex}.  If this is the correct explanation
for the extra strength in these neutrino interactions, then it could also have a significant
effect on the reconstructed neutrino energy in oscillation events, $E_\nu^{QE}$, which assumes quasielastic 
scattering from single nucleons within carbon~\cite{Martini:2012fa}.  In short, the reconstructed 
neutrino energy may be incorrect in a large fraction of the oscillation events leading  to incorrect conclusions
about the resulting fits to oscillation models.

The addition of scintillator will allow this idea to be experimentally tested.  With the 
scintillator addition proposed here,
the detector response to final state nucleons in a typical CCQE event will be increased by
about a factor of five.  This scintillation light is a measure of the total energy in the 
event ($E_\nu^{total}$) as opposed to that reconstructed from the just the lepton track, $E_\nu^{QE}$.
A comparison of $E_\nu^{QE}$ with $E_\nu^{total}$ will allow further insight into the two-nucleon
correlation issue in general and, specifically, into its relevance to the low-energy oscillation excess. 

\clearpage
\section{Increasing scintillation in MiniBooNE}
\label{sec:mb_scint_ADD}
In order to execute the main goal of a NC/CC test of the oscillation 
excess, the detector light output response to neutron captures in MiniBooNE must 
be increased.  This can be accomplished by adding a few hundred parts per million 
of a scintillator such as butyl-PBD or PPO to the MiniBooNE mineral oil.
The resulting level of scintillation will provide a distinct signature of 
neutron capture.
This technique was demonstrated and used by the LSND 
experiment~\cite{Athanassopoulos:1996ds} to search for $\nueb$ 
appearance oscillations via inverse $\beta$-decay with coincident neutron signal.

When a fast neutron of up to a few hundred MeV is produced by a neutrino 
interaction in mineral oil (CH$_2$), it rapidly loses energy by elastic and 
inelastic collisions with the mineral oil until it is reduced to thermal 
energies (i.e.~``thermalizes'').
This thermal neutron will capture on hydrogen 99.5\% of the time and 
on carbon the remaining times with a total characteristic lifetime 
of 186~$\mu$s for mineral oil.
The hydrogen capture $(n p \rightarrow d \gamma)$ yields a 
2.2~MeV $\gamma$, while the $^{12}$C capture emits a 
5~MeV $\gamma$.

In addition to allowing for neutron identification (i.e.~``n-tagging''), increasing the
scintillation light in MiniBooNE will also allow for better sensitivity to the sub-Cerenkov 
threshold events (e.g.~nucleons) in addition to the super-Cerenkov leptons.  

\subsection{Neutron capture}
\label{sec:lsnd_ncap}
By measuring the energy, position, and time of the event, a selection 
criteria can be formed which will select n-capture events that are 
correlated with the primary neutrino event with high efficiency yet reject 
accidental events.
The LSND experiment used the n-capture technique to select 
$\nueb p \rightarrow e^{+} n$ events via detection of the prompt $e^{+}$ 
signal in coincidence with the delayed 2.2~MeV photon from neutron
capture.
A neutron likelihood can be formed with the 
PMT hit multiplicity, prompt-delayed capture distance, and timing distribution, 
and it was able to achieve a 51\% detection efficiency for correlated n-capture 
events while only being susceptible to a 1.2\% accidental probability.  

The average PMT multiplicity from a 2.2~MeV $\gamma$ was 35 PMT hits in the LSND
experiment.  
As MiniBooNE and LSND are comparably sized detectors with similar readout PMTs, 
the same efficiency and accidental rate suppression can be achieved by raising the 
MiniBooNE scintillation to yield the same PMT multiplicity for 2.2~MeV $\gamma$.  
LSND, with 25\% photocathode coverage, was able to achieve this light
yield with 30~mg of butyl-PBD per liter of mineral oil.  
MiniBooNE has an 11\% photocathode coverage, and this requires a
higher amount of scintillator to achieve similar performance.

\subsection{Scintillation light in MiniBooNE, currently}
\label{sec:mb_scint_NOW}
Sub-Cerenkov-threshold particles can produce some light in the current MiniBooNE configuration
with ``pure'' mineral oil.  
This occurs because of aromatic benzene-like impurities in the MiniBooNE mineral oil, 
and the light output from these has been characterized in a variety of benchtop and 
in-situ tests~\cite{AguilarArevalo:2008qa, Raaf:2002mt, Brown:2004uy}.   
The various tests are in reasonable agreement and have enabled the scintillation 
and related fluorescence to be characterized and coded into the MiniBooNE detector 
Monte Carlo (MC) simulation of the optical model.  
The main result that can be used to benchmark the well-characterized MiniBooNE oil is
that it has a scintillation strength of 32 visible photons emitted per MeV of energy deposit 
at a wavelength peak of 350~nm, with width $\sigma=25$~nm, and a decay time 
of 34~ns~\cite{Brown:2004uy}.  

\begin{figure}
\centering
\fbox{\includegraphics[width=0.6\textwidth]{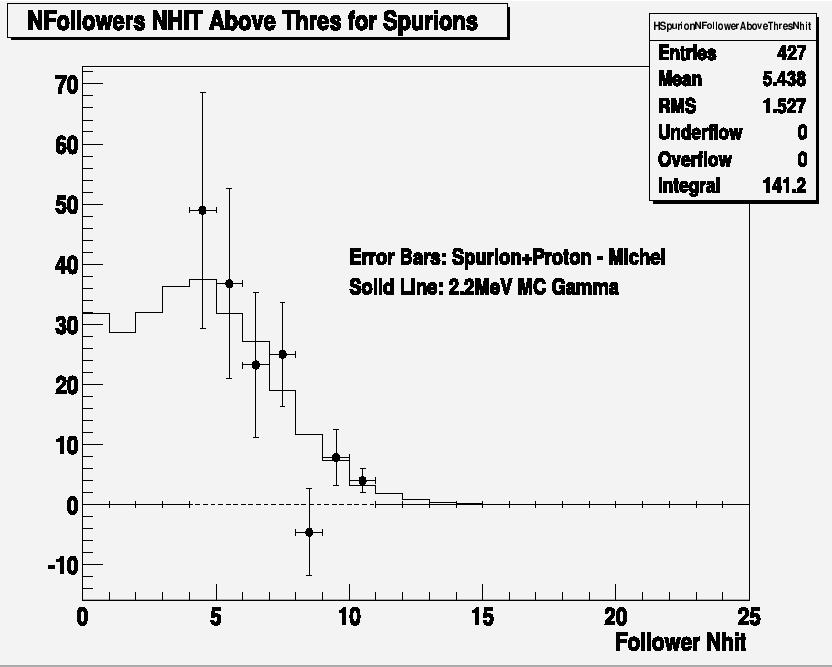}}
\caption{
PMT multiplicity distribution from neutron-capture candidate events in an 2004 data 
study~\cite{mbtn116}.  
The data points are shown with error bars after subtraction of random coincidence rate. 
There is no data below 4 hits because of the trigger threshold.  
The MC prediction is shown as solid histogram. 
``Spurion'' was a whimsical name given to neutron-capture candidates.}
\label{fig:tn116_nhit}
\end{figure}

The results of these scintillation tests have been folded into a complete,
optical-photon model that accounts for absorption, re-emission, and various scattering 
processes, and well-describes the MiniBooNE data.
The optical-photon model has been used to predict the current MiniBooNE response 
(undoped with additional scintillators) to 2.2~MeV $\gamma$, and it predicts a
multiplicity of approximately 5 PMT hits with about 80\% of the light in the directed Cerenkov cone.
The remaining 20\% of light is due to both isotropic scintillation and the isotropic re-emission 
of scattered UV Cerenkov light.
These results have been verified in a study with MiniBooNE data in 2004~\cite{mbtn116} 
to determine the feasibility of adding a neutron-capture trigger 
at that time. 
Figure~\ref{fig:tn116_nhit} shows the PMT multiplicity distribution for neutron-capture 
candidate events (``follower'' events) derived from this study.   
This shows that the MC correctly simulates the current level of scintillation and response 
to 2.2~MeV n-capture events, and that the response is too low to conduct a neutron search 
with the current scintillation level (nominally undoped, less the fluorescent impurities).  

\subsection{Simulations of increased scintillation light}
\label{sec:sim_scint}
We first assess how much light must be produced by a given scintillator before determining 
the amount of scintillator required.  
Using the MiniBooNE detector MC, the total scintillation fraction can be increased, 
and the PMT multiplicity simulated for 2.2~MeV n-capture $\gamma$.  
As the scintillation level is increased by a scaling factor $k$, the total isotropic response 
(not in the prompt, directed Cerenkov light cone) increases faster than $k$ because 
the scintillator absorbs Cerenkov light in the UV (not detected by the PMTs) and shifts it into the 
PMT wavelength sensitivity range.  
This additional contribution is isotropically re-emitted and is then detected by the PMTs.  

\begin{figure}
\centering
{\includegraphics[width=0.7\textwidth]{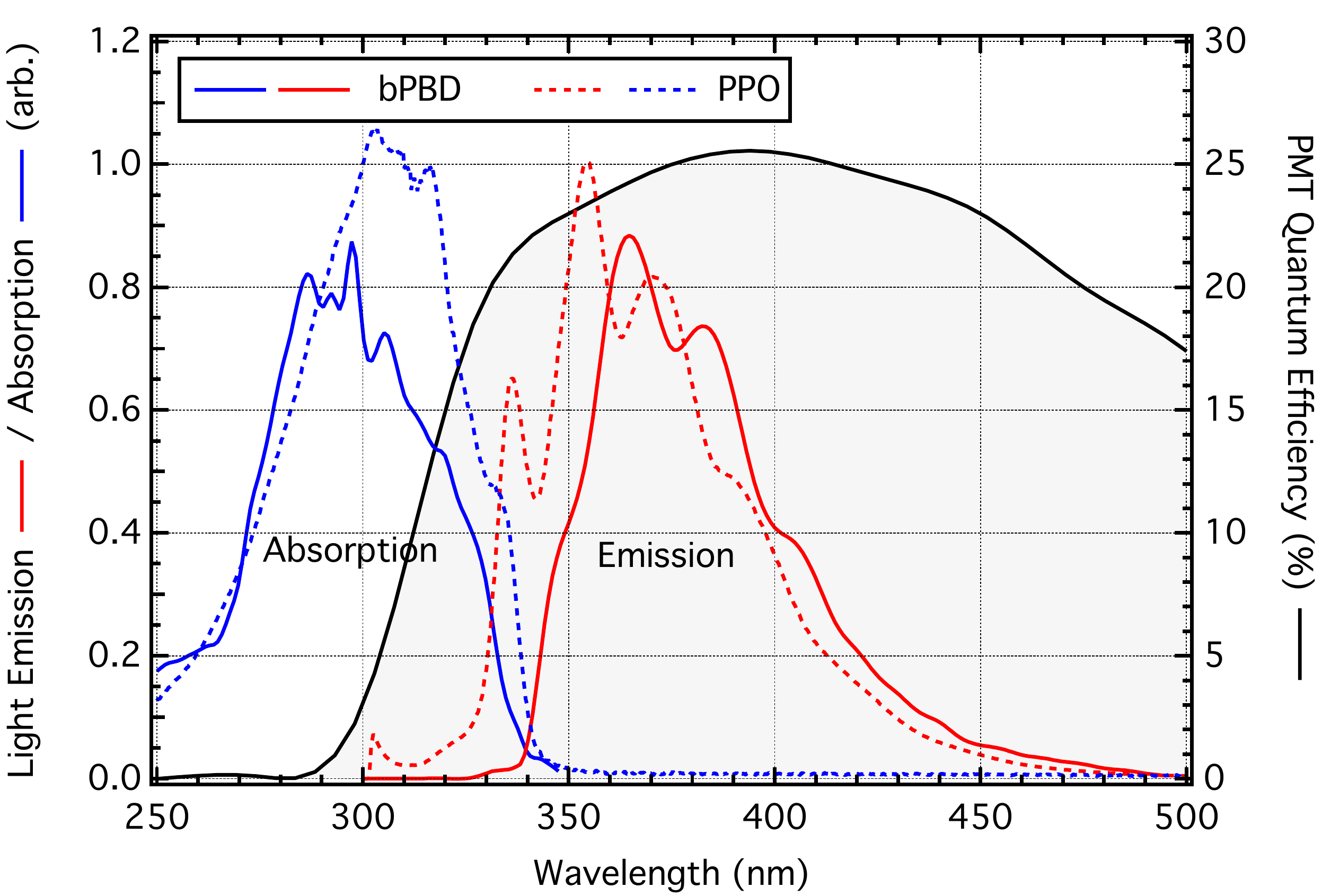}}
\caption{The absorption and emission spectra of butyl-PBD and PPO with the MiniBooNE
PMT quantum efficiency overlaid on the right.}
\label{fig:em_ab_vs_QE}
\end{figure}

Initial light output tests were performed with a naive scintillation model by scaling 
the 350~nm light emitted by the current fluors present in the MiniBooNE oil.  
A more sophisticated model was developed that utilized a fluor model more appropriate 
for butyl-PPD, and this model was based upon the emission and absorption spectra for 
butyl-PBD which are shown with the PMT quantum efficiency (QE) in 
Fig.~\ref{fig:em_ab_vs_QE}.  
A fluor such a butyl-PBD is more efficient at increasing the scintillation compared to 
the current fluors in the MiniBooNE oil because the emission spectrum has a longer 
average wavelength for which the attenuation length of the mineral oil is longer 
and the PMT QE is higher.  
Also, butyl-PBD has comparable fast and slow scintillation components 
(2~ns and 20~ns, respectively) in contrast to the current MiniBooNE fluors 
which contain just a slow component (34~ns).  
The fast scintillation component significantly improves the reconstruction of the
n-capture position reconstruction as will be seen below in Sec.~\ref{sec:event_recon}.

\begin{figure}
\centering
{\includegraphics[width=0.6\textwidth,trim=8mm 10mm 5mm 5mm, clip]{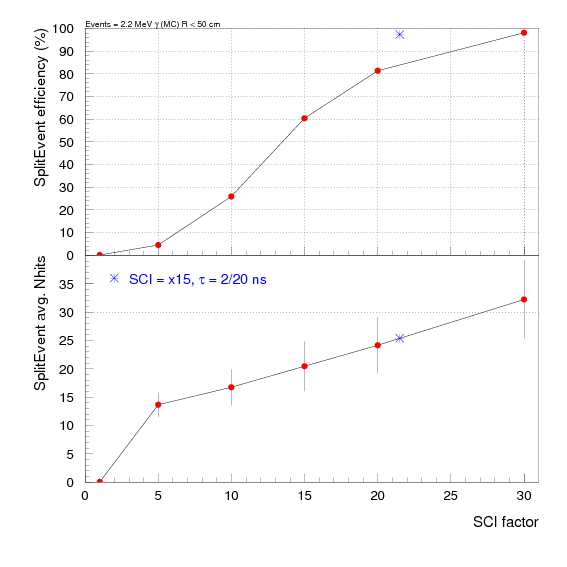}}
\caption{Neutron capture (2.2~MeV $\gamma$) event detection efficiency (top) and 
the average PMT multiplicity distribution (bottom) is shown as a function of 
scintillation strength multiplier (SCI factor).   
The red points are from the naive scintillation model where the current 350~nm MiniBooNE 
mineral oil emission is scaled from current value.  
The blue point is for the butyl-PBD model at scintillation strength of 15 times the
current MiniBooNE mineral oil and with light output equivalent to the naive scintillation
model with a strength multiplier of about 22.}
\label{fig:nhse_vs_sci}
\end{figure}

The results from the MC studies are summarized in Fig.~\ref{fig:nhse_vs_sci}.  
Note that the efficiency for finding events over threshold (PMT multiplicity) of 10 hits 
increases with the scintillation multiplier as does the average PMT multiplicity.  
A scintillation multiplier of 1 represent no additional scintillation beyond that of the
current MiniBooNE fluors in mineral oil (32 photons\,/\,MeV).  
Also note that the butyl-PBD model with a scintillation multiplier of 15 is equivalent 
to the naive model at about a value of 22 because butyl-PBD has a more optimal 
overlap of its emission spectrum with the PMT QE and there is a longer attenuation length 
of mineral oil at these wavelengths.
The detection efficiency and PMT multiplicity results for butyl-PBD lead to the 
conclusion that a multiplication factor of 15 times the current MiniBooNE value 
is sufficient to reconstruct the 2.2~MeV n-capture $\gamma$.  

\subsection{Event reconstruction}
\label{sec:event_recon}
Based on the MC studies of increased scintillation light in n-capture events, an 
increase of the scintillation strength by a factor of 15 with butyl-PBD (or similar)
should be adequate.  Further studies were performed using the existing MiniBooNE 
reconstruction apparatus to determine how well 2.2~MeV $\gamma$  and higher energy
electrons are reconstructed with their increased scintillation light.

\begin{figure}
\centering
{\includegraphics[width=0.7\textwidth,trim=8mm 10mm 5mm 5mm, clip]{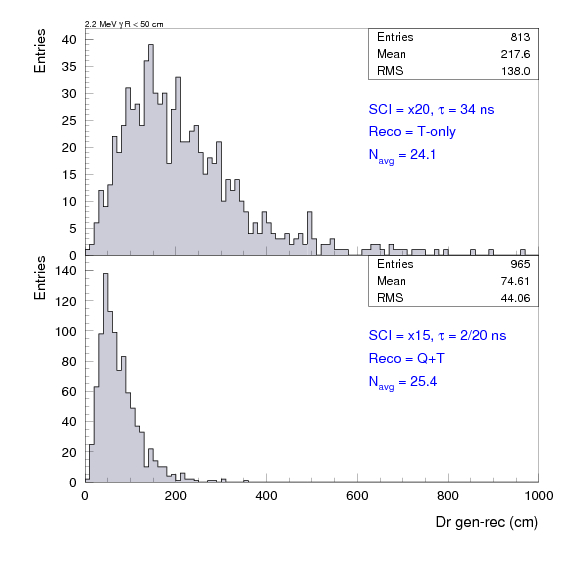}}
\caption{Position resolution of simulated 2.2~MeV $\gamma$ events for the naive scintillation
model (top) compared to that for the butyl-PBD model (bottom).}
\label{fig:dr_22g}
\end{figure}

Low-energy events such as 2.2~MeV $\gamma$  are reconstructed using PMT charge and time
information.  For position reconstruction, the time information is the most critical
as the time of arrival at each tube allows the position to be determined.  It is
crucial to have a scintillator with a fast time component or the position information
is degraded.  This is especially true with low-light events where the PMTs are typically
detecting an average of one photoelectron.  These effects can be seen in Fig.~\ref{fig:dr_22g} where
the position resolution for 2.2~MeV $\gamma$  is plotted for both the naive and butyl-PDB scintillation
models,  using new reconstruction algorithms developed specifically for this task.
The butyl-PBD model with a fast scintillation component of 2~ns results in improved
position reconstruction of $\approx$75~cm.  This is comparable to that achieved in
the LSND n-capture reconstruction and, therefore, should yield similar n-capture efficiency
and accidental background rejection.

\begin{figure}
\centering
{\includegraphics[width=0.8\textwidth,trim=8mm 92mm 5mm 5mm, clip]{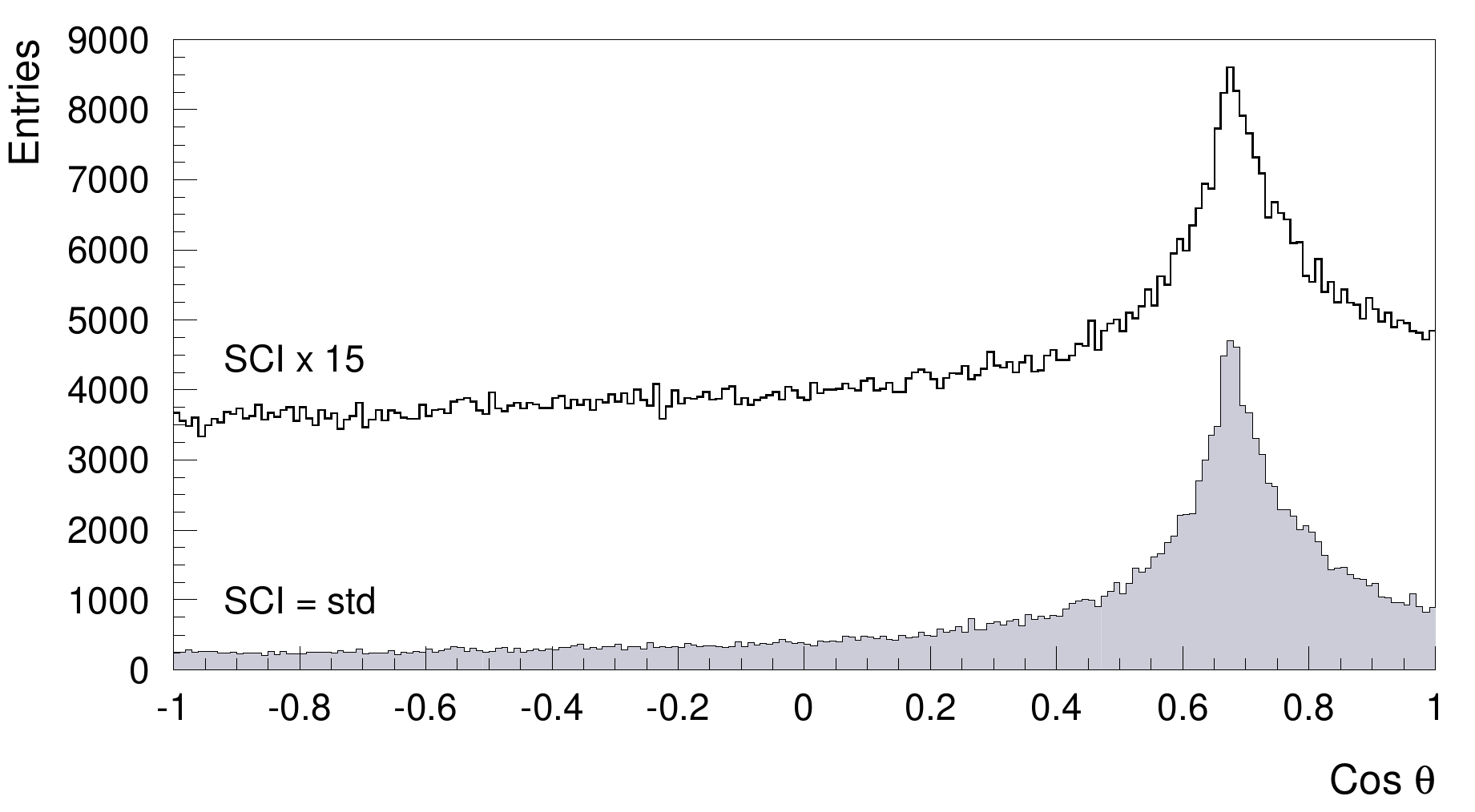}}
\caption{Angular distribution of light around the reconstructed 
direction of 50~MeV electrons for the standard (current) scintillation model compared to the butyl-PBD model.}
\label{fig:q_angular_e50mev}
\end{figure}

Adequate reconstruction of n-capture events is only part of the reconstruction battle.  This
increased level of scintillation light must not degrade the electron reconstruction which
is crucial for an oscillation analysis.  This was studied by generating electron events using 
the detector MC with the butyl-PBD scintillation model and then reconstructing these events using
the current MiniBooNE algorithms.  This was first tested with 50~MeV electrons.  The Cerenkov 
cone is still readily located within the increased
scintillation light as can be seen in Fig.~\ref{fig:q_angular_e50mev} where the angular distribution 
of the light around the reconstructed electron direction is plotted. The level of isotropic light
is increased by about a factor of 15 as expected, but the higher level of directed light in the 
Cerenkov cone is still quite obvious.  

The reconstruction of higher energy electrons has also been studied.  In order to
obtain the optimum reconstruction of neutrino events, the PMT charge and time likelihoods will
need to be redetermined for the new scintillator mixtures.  This has not been addressed fully
at this time. However, a few steps in that direction have been taken and these intermediate
results allow some conclusions to be made. First some details need to be explained.

\begin{figure}
\centering
{\includegraphics[width=0.9\textwidth,trim=8mm 12mm 7mm 7mm, clip]{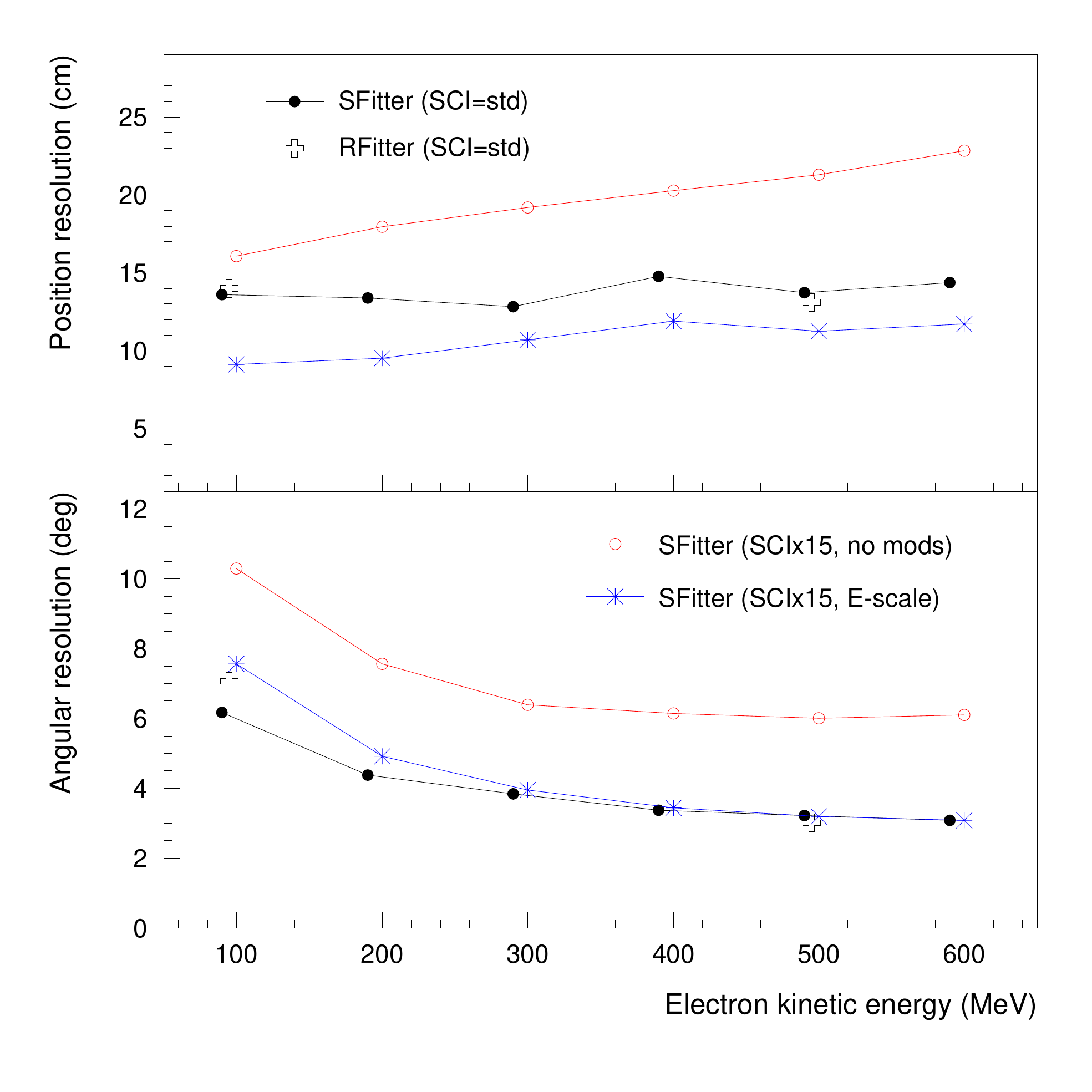}}
\caption{Position (top) and angular (bottom) resolution obtained with the MiniBooNE
reconstruction ``SFitter'' and ``RFitter'' algorithms for 100-600~MeV electrons 
with both the standard (current) scintillation model and a butyl-PBD model. The Rfitter was
run at only 100 and 500~MeV.}
\label{fig:reco_el_vs_e}
\end{figure}

MiniBooNE has several levels of reconstruction, the first ``SFitter'' assumes a point like
model for the lepton track.  The second ``RFitter'' contains an extended track 
model~\cite{Patterson:2009ki} and was used for the particle identification (PID) in the 
main oscillation analysis but is very sensitive to tuning which has not yet been performed.
The SFitter has a more complex PID algorithm but is  less sensitive to tuning.   
Results from applying these reconstruction algorithms 
to the simulated 100-600~MeV electrons are shown in Fig.~\ref{fig:reco_el_vs_e}.  The SFitter 
and RFitter results are shown with the standard (current) scintillation model.  The SFitter
was run with no modifications to the underlying likelihoods and then with only a minor adjustment
to the energy scale.  There are a few things to note in this plot.  First, the untuned SFitter
with increased scintillation returns degraded resolution values across the energy range.
This is expected as the underlying parameters of the fit are incorrect in this case.
However, the SFitter with minor modifications actually has better position resolution than
the standard case.  It approaches the standard value at higher energy for angular resolution.
The performance of the untuned RFitter with increased scintillation is off the scale in
 Fig.~\ref{fig:reco_el_vs_e}.

The conclusions from these studies at this time are that simple changes to the SFitter
recover resolution to the standard level after a minor change to adjust for the stronger scintillation
light.   We expect to improve this for the SFitter with proper tuning of the time likelihoods 
and underlying optical model.  From
the behavior of the SFitter, we expect to recover the good resolution for the RFitter after
the required retuning of the algorithm.

The results of these studies indicate that an increase of scintillation light
with a butyl-PBD type dopant by about a factor of 15 is sufficient for the desired
n-capture reconstruction accuracy.  In the studies to date, this level of increased
scintillation light indicates that the electron reconstruction should not suffer
-- at least for energies below 600~MeV.

If electrons can be reconstructed sufficiently, it seems reasonable to assume that
$\pi^0$s and $\mu$s may also be reconstructed with accuracy as in existing analyses.
However, that has not yet been shown conclusively and is subject of current work.
It is an important question.

\subsection{Determination of the scintillator cocktail}
\label{sec:scint_mix}
From the studies in the preceding sections, we have determined that we
should add an amount of scintillator with absorption/emission spectrum and 
lifetime of butyl-PBD to obtain a light yield of 15 times that the current level in MiniBooNE.  

\begin{figure}
\centering
{\includegraphics[width=0.8\textwidth]{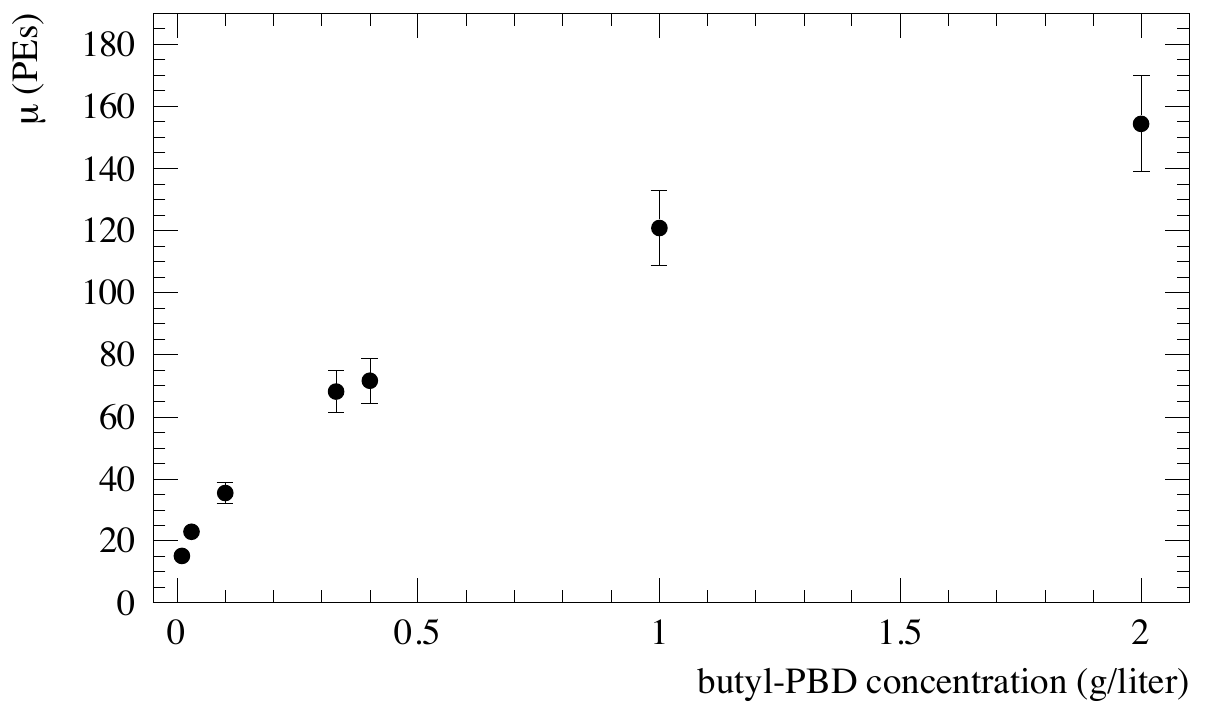}}
\caption{MiniBooNE tests on light level vs.~concentration of butyl-PBD~\cite{mbtn74}.  
The vertical scale is number of photoelectrons (PE) produced with that particular apparatus.}
\label{fig:PBDvsConc_TN74}
\end{figure}

To determine the scintillator concentration that will produce this light level,
we have examined data collected during MiniBooNE oil tests in 2001-2002 together
with recent tests.
In 2001-2002, the level of scintillation in candidate mineral oils for MiniBooNE
was measured at the Indiana University Cyclotron Facility proton beam~\cite{mbtn74}.
This beam was ideal for these tests as the 200~MeV proton beam is below the 
Cerenkov production threshold, and a direct measurement of the isotropic scintillation 
light could be isolated.  
While the tests focused on the scintillation from pure mineral oil, several scintillators 
were tested with the idea that a slight amount of scintillator could be beneficial for
MiniBooNE particle identification.  
The tests with butyl-PBD are shown in Fig~\ref{fig:PBDvsConc_TN74}.  
In separate tests, it was determined that scintillation from pure MiniBooNE mineral oil 
produced 5 photoelectrons with the same experimental setup.  
Therefore, in order to produce a factor of 15 more light than pure MiniBooNE mineral 
oil (Sec.~\ref{sec:sim_scint}), approximately 0.3~g\,/\,l of butyl-PBD is desired.  
Similar tests of butyl-PBD were carried out in service of LSND~\cite{Reeder:1993ff} and
are consistent with the MiniBooNE results.  

In the process of considering 
butyl-PBD\footnote{Butyl-PBD is a synonym for 2-(4-tert-Butylphenyl)-5-(4-biphenylyl)-1,3,4-oxadiazole,
CAS Number 15082-28-7}  as a candidate scintillator, 
PPO\footnote{PPO is a synonym for 2,5-Diphenyloxazole, CAS Number 92-71-7} 
was identified as another possibility.  
As can be seen in Fig.~\ref{fig:em_ab_vs_QE} the absorption / emission
spectra are quite similar between butyl-PBD and PPO.  
In addition, the light output at a given concentration of PPO in mineral oil is about 
the same as butyl-PBD, and the characteristic emission times are about the same 
with a fast component around 2~ns~\cite{Birks:1964zz}.
In preliminary investigations about price, PPO looks to be about 1/4 of the price
per kg of butyl-PBD.

\begin{figure}
\centering
{\includegraphics[width=0.7\textwidth]{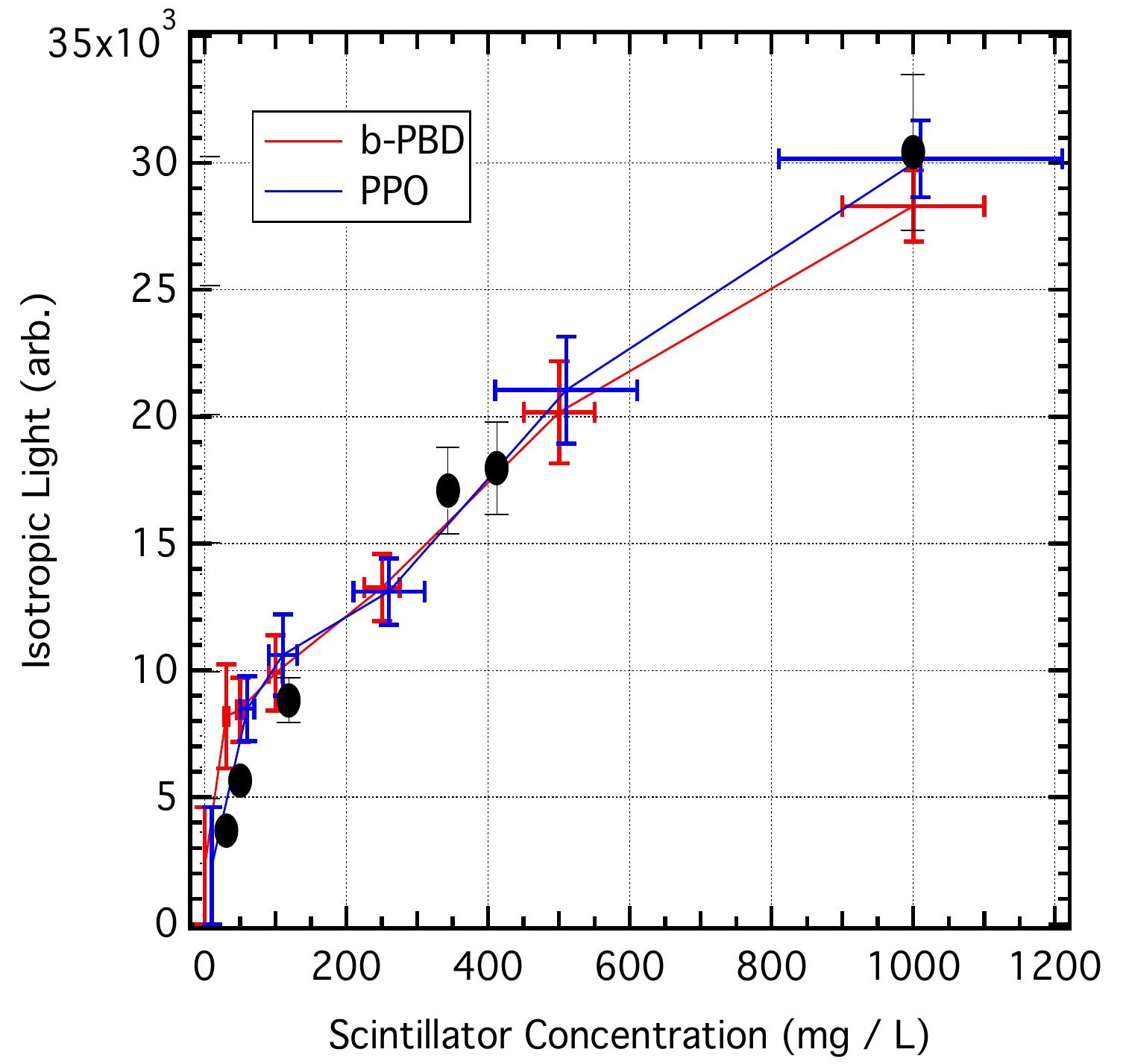}}
\caption{Recent tests of isotropic light output from butyl-PBD and PPO in mineral oil 
as a function of scintillator concentration.  
The recent lab tests are shown as red and blue points.  
The data taken from early mineral oil tests with butyl-PBD~\cite{mbtn74} are shown as
 black points and are overlaid.}
\label{fig:LightOut}
\end{figure}

In order to investigate PPO further and to verify the old results on butyl-PBD, 
the test setup from the 2001-2002 IUCF tests was modified.  
Instead of operating in the 200~MeV proton beam, the test setup utilized
a strong, collimated $^{60}$Co $\gamma$ source with the original oil test apparatus.  
The light output of various concentrations of butyl-PBD and PPO in mineral oil were tested,
and the results are shown in Fig.~\ref{fig:LightOut}.
From these tests it was determined that the light output from butyl-PBD and PPO at
a given concentration are the same.  
The data from 2001-2002 IUCF tests~\cite{mbtn74} on butyl-PBD are overlaid on the plot 
with an arbitrary matching of scale as the light collection efficiency in the current setup 
is not absolutely calibrated.  
However, the behavior with scintillator concentration is the same between the different tests.  
If the old data is used to set the scale, then our current tests indicate that PPO has the same 
light output as butyl-PPD and a concentration of 0.3~g/l of PPO will yield a factor of 15 light
increase in the MiniBooNE mineral oil.

\clearpage
\section{Simulated oscillation analysis}
\label{sec:analysis}
In this section we report the results of a simulated analysis to determine the
sensitivity of an oscillation search with NC/CC identification via identification
of n-capture events. The method assumes that the oscillation search will be repeated
with a new data set, after adding scintillator to MiniBooNE.  It also assumes
that a candidate oscillation sample is obtained using the same particle ID strategy
and, for this study, that the backgrounds and excess are the same as obtained from
the current MiniBooNE results.  The neutron-capture analysis is then performed
with this simulated oscillation sample.

\subsection{Measuring the neutron fraction in oscillation candidate events}
\label{sec:NC_CC_study}

\begin{figure}
\centering
{\includegraphics[height=0.83\textheight]{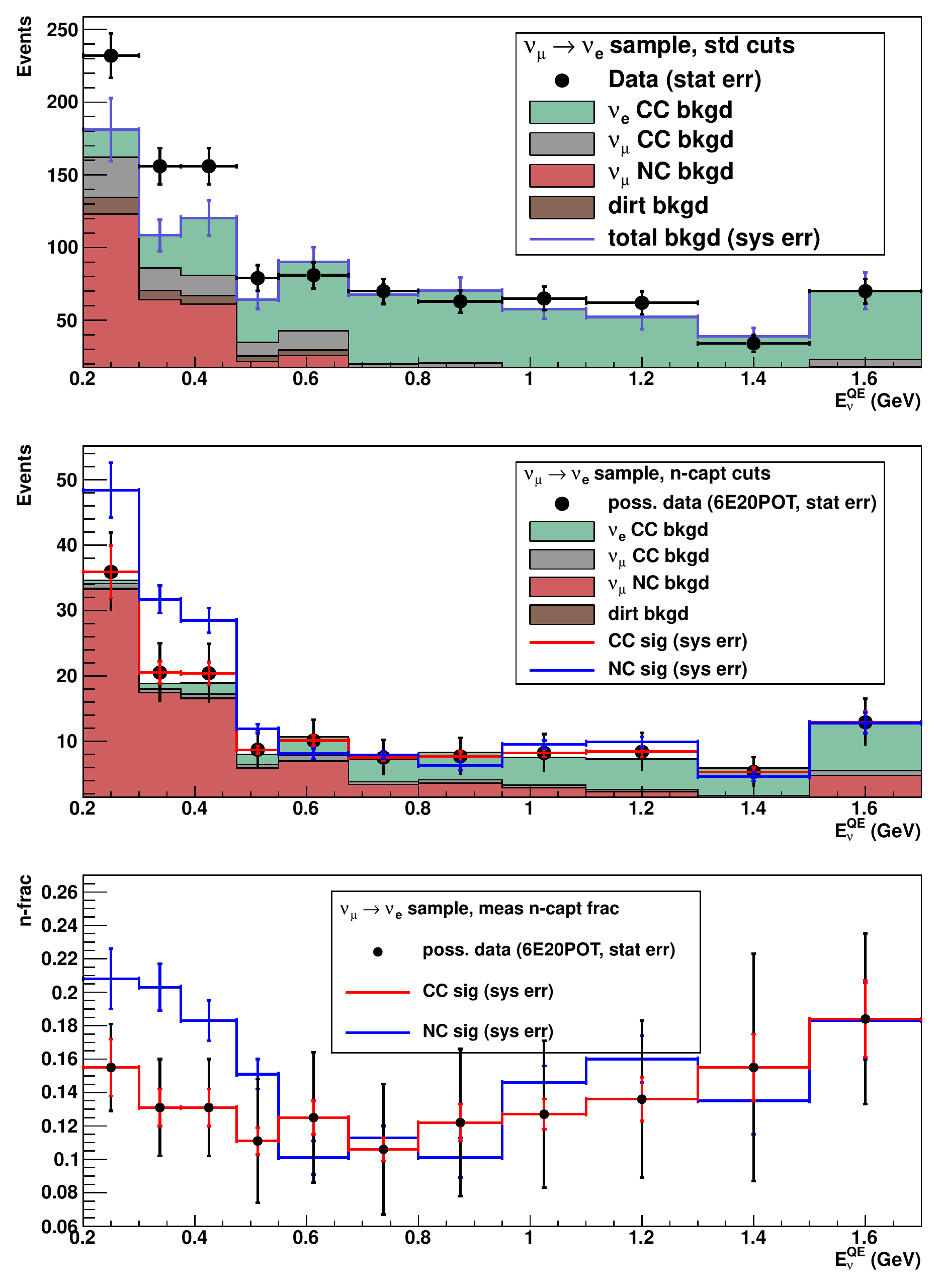}}
\caption{Simulated neutron fraction measurement using the existing MiniBooNE neutrino
data set. The $E_\nu^{QE}$ distribution of the current data set is shown in the top plot, 
the middle plot contains events that have an associated n-capture event, and 
bottom is the simulated measured neutron fraction.  In the middle and top plots, the 
red (blue) curve is the prediction if the excess is due to CC signal (NC background) with
systematic errors and the data points are drawn using the CC signal hypothesis with 
statistical errors.
}
\label{fig:ncapt_ev}
\end{figure}

A simulated measurement of the neutron fraction in oscillation candidate events
was performed using the current MiniBooNE neutrino oscillation data set and is summarized
in Fig.~\ref{fig:ncapt_ev}.  Figure~\ref{fig:ncapt_ev}(top) shows the $E_\nu^{QE}$ 
distribution of the simulated data events along with the background predictions and is the 
same as Fig.~\ref{fig:nu_nubar_stack}(top) excepting that unnormalized events plotted are shown
and the backgrounds have been regrouped into four categories: $\nue$ CC, $\numu$ CC, $\numu$ NC,
and dirt.  This study assumes that the systematic and statistical errors will be the same as in the  
the existing MiniBooNE neutrino set which resulted from $6.5\times10^{20}$~POT.
 
For this study, each of the four backgrounds were modeled to predict
the fraction (before n-capture efficiency) of events with an associated neutron.
For the $\numu$ CC and $\nue$ CC background events, the model assumed that 1\% of
events in the 250~MeV bin had associated neutrons and rising to 10\% at 1~GeV.  This
low energy behavior is based on a measurement by the LSND experiment~\cite{Athanassopoulos:1997rn} 
of the neutron fraction in  $\numu~^{12}C \rightarrow \mu^{-} X$ events at mean $\numu$ energy 
above $\mu^{-}$ threshold of 156~MeV.  The upper limit comes from an 
$^{12}C(e,e^{\prime} p)X$  experiment~\cite{Garino:1992ca} that measured nuclear transparency
at an average proton recoil energy of 180~MeV. Note that this model, since it is based on 
actual measurements from carbon, includes the effects of final state interactions of
the nucleons as the exit the nucleus.

For the $\numu$ NC events, a neutron fraction of 50\% was assumed.  This is based on 
models that predict the dominance of incoherent $\Delta$ production in  
both NC$\pi^{0}$ and NC$\gamma$ production~\cite{Zhang:2012ak,Hill:2009ek} and 
that $\Delta$ will decay symmetrically to neutron or proton.  The ``dirt'' background is
assumed to be neutron free since it is likely photons that are produced in the material
around the MiniBooNE detector tank and escape detection as they travel through the veto.
This assumption is not critical since the dirt background is a small fraction of sample.
The neutron fraction of the dirt background will be measured in the actual analysis.
For the  $\nue$ CC oscillation signal, the neutron fraction will be the same as for
the $\nue$ CC background described above. 

These assumptions will need to be fine tuned with a more complete model.
The recent work of Refs.~\cite{Serot:2011yx}-\cite{Hill:2009ek} together with actual
measurements of the reaction channels contributing to the backgrounds (Sec.~\ref{sec:n-calib}) will
make this possible.  The central value of the predicted neutron fraction will actually 
be somewhat different than 50\% due to NC scattering from free protons (in CH$_2$) and coherent
scattering which do not result in neutrons in the final state. However, final-state interactions
of neutrons in NC interactions producing additional neutrons, will move the neutron fraction
to higher values. These uncertainties are explored by  varying the underlying assumptions 
and is also quantified in following paragraphs.

In addition to a model for the fraction of neutrons in the signal and background channels,
a neutron efficiency was needed as input.  This was set to be the same as for the LSND
experiment, 50\%~\cite{Aguilar:2001ty}, since the detection method and scintillation light 
level will be similar. One other input required for this study is the probability for accidental
neutrons passing the n-capture cuts.  This was set to be 2\%, again the same as for the LSND
experiment.

The systematic errors identified and quantified in this study are from the following
uncertainties:
\begin{itemize}  
\item Predicted NC, CC, and dirt backgrounds.  The values for the associated uncertainties 
are extracted from the MiniBooNE constrained oscillation fit and are 10-12\% in the lowest energy bins
where the NC backgrounds dominate.
\item Predicted NC and CC neutron fractions. These errors are set to 5\% relative for both NC and CC 
channels. 
\item Assumed n-capture efficiency.  Set to 1\% uncertainty as it will be the same value and 
completely correlated between the background and signal assumptions.
\end{itemize}

When these simulated n-capture cuts are applied to the study sample of Fig.~\ref{fig:ncapt_ev}(top),
the event distribution shown in Fig.~\ref{fig:ncapt_ev}(middle) results.  Since the CC
backgrounds contain very few neutrons they are greatly reduced in the resulting sample.
The n-capture cuts are effectively an NC selection.  The oscillation
excess, assumed in this study to be all $\nue$ CC oscillation events, disappears with
the neutron requirement.
This can be seen as the simulated data distribution is the same as the prediction with
the CC signal assumption.  On the other hand, if the oscillation excess is instead due to
an incorrect calculation of the NC background, the sample will instead show an excess as shown 
for the NC ``signal'' in the figure.

This situation is more easily examined in Fig.~\ref{fig:ncapt_ev}(bottom) where the 
neutron-fraction distribution is plotted. Again, the data is assumed to have an excess
over background due to $\nue$ CC oscillation events so it overlays the CC signal prediction.
The NC prediction is significantly above the CC prediction and the two possibilities
are quite distinct in the lowest $E_\nu^{QE}$ where the excess appears.  In this plot,
the total estimated systematic errors are shown on the predictions and the resulting
statistical errors on a $6.5\times10^{20}$ POT sample are shown on the simulated data points.

Using this simulated neutron fraction, an oscillation excess significance
test was extracted by calculating the separation between the simulated data with
the excess assumed to be $\nue$ CC oscillation events (the data points with statistical
errors in Fig.~\ref{fig:ncapt_ev}(bottom)) and the NC prediction (the top histogram
in Fig.~\ref{fig:ncapt_ev}(bottom) with systematic errors). The resulting significance
is 3.5$\sigma$ and is limited by statistical error. It has effectively no correlation
with the first-stage oscillation result (without n-capture).  So, the 3.5$\sigma$ excess can
be combined \textit{independently} with the 3.4$\sigma$ neutrino-mode excess (if it is present in
this new data set).  The combination yields a significance of  $\approx5\sigma$.

Another method to quantify these simulated results is a null exclusion test.
This was performed by forming a $\chi^{2}$ from the simulated data as 
compared to the null hypothesis (NC prediction).  The sum in this test ran
over all $E_\nu^{QE}$ in the neutron-fraction distribution of 
Fig.~\ref{fig:ncapt_ev}(bottom).  The resulting null exclusion probability, if
the excess is due to $\nue$ oscillations, is 0.03\%.

\subsection{Variations of study assumptions}
The assumptions explained in the previous section were varied in alternate 
models in order to determine the effect on the resulting significance of the
oscillation hypothesis test.  These studies are summarized in Table~\ref{tab:nfrac_sig}.

\begin{table}
\centering
\begin{tabular}{|l|rrr|rrr|rrr|c|}
\hline
                                & \multicolumn{6}{|c|}{neutron fraction} &  & & & \\ \hline
configuration                   & \multicolumn{3}{|c|}{NC prediction}    & \multicolumn{3}{|c|}{fake data}  & 
\multicolumn{3}{|c|}{difference} & n$\sigma$ \\ \hline \hline
\textbf{standard}                   & 0.191 & $\pm$ & 0.008 & 0.134 & $\pm$ & 0.015 & 0.057 & $\pm$ & 0.016 & \textbf{3.48} \\
4E20POT                             & 0.191 & $\pm$ & 0.008 & 0.134 & $\pm$ & 0.018 & 0.057 & $\pm$ & 0.019 & 2.95 \\
2E20POT                             & 0.191 & $\pm$ & 0.008 & 0.134 & $\pm$ & 0.026 & 0.057 & $\pm$ & 0.027 & 2.16 \\
(bckgnd error)$\times$0.5           & 0.191 & $\pm$ & 0.005 & 0.134 & $\pm$ & 0.015 & 0.057 & $\pm$ & 0.015 & 3.73 \\
(n-capture efficiency)=0.75         & 0.277 & $\pm$ & 0.012 & 0.191 & $\pm$ & 0.018 & 0.086 & $\pm$ & 0.021 & 4.13 \\
(accidental efficiency)$\times$2    & 0.211 & $\pm$ & 0.008 & 0.154 & $\pm$ & 0.016 & 0.057 & $\pm$ & 0.017 & 3.29 \\
(CC n-fraction)$\times$2            & 0.191 & $\pm$ & 0.008 & 0.137 & $\pm$ & 0.015 & 0.054 & $\pm$ & 0.017 & 3.26 \\
(low-E CC n-fraction)=0.06          & 0.199 & $\pm$ & 0.008 & 0.147 & $\pm$ & 0.015 & 0.051 & $\pm$ & 0.017 & 3.00 \\
(NC n-fraction error)$\times$2      & 0.191 & $\pm$ & 0.010 & 0.134 & $\pm$ & 0.015 & 0.057 & $\pm$ & 0.017 & 3.31 \\
dirt n-fraction=0.5                 & 0.203 & $\pm$ & 0.008 & 0.145 & $\pm$ & 0.015 & 0.057 & $\pm$ & 0.017 & 3.32 \\
(NC bckgnd)$\times$2                & 0.215 & $\pm$ & 0.011 & 0.175 & $\pm$ & 0.014 & 0.040 & $\pm$ & 0.017 & 2.29 \\
(NC bckgnd)$\times$2 + $\infty$ POT & 0.215 & $\pm$ & 0.011 & 0.175 & $\pm$ & 0.000 & 0.040 & $\pm$ & 0.010 & 3.81 \\
(NC n-fraction)= 0.42               & 0.165 & $\pm$ & 0.006 & 0.117 & $\pm$ & 0.014 & 0.048 & $\pm$ & 0.015 & 3.17 \\
$\infty$ POT                        & 0.191 & $\pm$ & 0.008 & 0.134 & $\pm$ & 0.000 & 0.057 & $\pm$ & 0.008 & 7.63 \\
\hline \hline
\end{tabular}

\caption{Summary of simulated results for a neutron fraction measurement with the oscillation sample.
In this study the ``NC prediction'' is the expected value for neutron fraction with systematic
errors in the case if the  oscillation excess was all from NC backgrounds.  
The ``fake data'' is with the assumption that the oscillation excess is all CC oscillation signal with
statistical errors.  The top row (``standard'') is with the standard assumptions as explained in 
Sec.~\ref{sec:NC_CC_study} and with $6.5\times10^{20}$~POT, and the other rows are variations with the standard 
configuration plus the change indicated.}
\label{tab:nfrac_sig}
\end{table}

The top row of Table~\ref{tab:nfrac_sig} summarizes the results of the significance
study with the assumptions explained in Sec.~\ref{sec:NC_CC_study}  which yields a
significance value of 3.5$\sigma$. Note that this includes the assumption of $6.5\times10^{20}$~POT. 
The subsequent rows show the values for some
possible variations in underlying assumptions.  There are a few values of note.  
If the data sample is reduced to $2\times10^{20}$~POT the significance drops to $2.2\sigma$.
This is indicative of the fact that this is a statistics limited test.  Note that
with infinite POT the significance goes to $>7\sigma$.  While infinite POT is not
possible, it points out that further optimization of cuts may be possible to increase
statistics while taking some hit in systematic errors.  This will be a subject of
future study.

Neglecting the event statistics variations, the significance ranges from approximately
2.3$\sigma$ to 4.1$\sigma$. The lowest value of 2.3$\sigma$ is from an increase in
the size of the NC background by a factor of 2. This may happen if the signal to background
ratio in the oscillation sample is not held at the current levels with the new analysis
after the addition of scintillator.  This is an important point. The addition of 
scintillator to MiniBooNE must not cause a significant change in the oscillation
signal purity. 

Another variation to note is that where the neutron fraction of NC events is changed
from 50\% to 42\%.  This is the value obtained with consideration of free protons. 
It does lower the significance slightly as does an increase in the assumed neutron fraction
of CC current events.  However, the significance in the cases does not drop below 3$\sigma$
and it is expected that consideration of neutron final state effects in carbon will increase this.
While these variations in the neutron fraction of events may occur it is important
to remember that this will not be an unknown variation. These fractions will be measured
in calibration samples as explained in the next section.

\subsection{Calibration of the neutron fraction in NC and CC events}
\label{sec:n-calib}
The MiniBooNE low energy excess has stimulated recent theoretical work in non-oscillation
explanations of the data, in particular, that the NC$\gamma$ background has been 
underestimated~\cite{Zhang:2012new}-\cite{Hill:2010zy}.
Indeed, if this background was not correctly estimated in the MiniBooNE oscillation analysis, it could
be the explanation of the excess.  However, recent theoretical results are actually supporting the
MiniBooNE calculations~\cite{Zhang:2012ak}.  A side benefit of this theoretical work is that realistic models
for NC processes at these energies ($E_\nu \approx 100-500$~MeV) have been improved and better 
understood which will help any future experiments working in this area. In addition,  these models
can be used to better understand the n-fraction in NC events which is crucial for the experiment
proposed here. However, an important advantage to the n-capture method is that the neutron fraction from both
CC and NC events can be measured with this experiment.  This, together with model guidance, 
will keep systematics on backgrounds quite low. 

For CC events, one wants to know, as a function of reconstructed neutrino and, in particular at
$E_\nu \approx 100-500$~MeV,  the neutron fraction in the reaction $\nu_e C \rightarrow e^{-} X$ 
(Fig.~\ref{fig:fd_CCNC_set:a}).  This has been measured at this energy scale by 
LSND~\cite{Athanassopoulos:1997rn} in the $\nu_\mu C \rightarrow \mu^{-} X$ reaction as described earlier.
The difference between $\nu_e$ and $\nu_\mu$ can be calculated with consideration of the higher
energy threshold in the $\nu_\mu$ reaction.  This will also be measured by MiniBooNE via 
$\nu_\mu C \rightarrow \mu^{-} X$ events that contain a neutron, identified with the n-capture
signal.  It is estimated (and used in the study above) that a 5\% systematic error may be obtained.

For NC events, the desired quantity is the neutron fraction for the NC oscillation backgrounds,
NC$\pi^0$ and NC$\gamma$ (Figs.~\ref{fig:fd_CCNC_set:b},~\ref{fig:fd_CCNC_set:c}).  The NC$\pi^0$
background can be measured via $\nu_\mu C \rightarrow \pi^0 X$ where the $\pi^0$ is correctly
identified in the detector.  This measurement has previously been 
made by MiniBooNE~\cite{AguilarArevalo:2008xs,AguilarArevalo:2009ww}.
The extrapolation to the background NC$\pi^0$ events is then performed with the detector simulation
and is fairly model independent, depending mainly on $\pi^0$ kinematics.

The NC$\gamma$ background may then be calculated from the radiative decay of the $\Delta$ with
the production rate of the $\Delta$ constrained by the NC$\pi^0$ measurement.   The change of
the  $\Delta$ width in carbon is taken into account.  This was the method used by MiniBooNE
in the current analysis and has been vindicated with separate theoretical 
calculations~\cite{Zhang:2012new,Hill:2010zy}.  It is important to note that if there is an
anomalous rate for NC$\gamma$, underestimated by both MiniBooNE and current theories, it is likely
to produce neutrons ~50\% of time.  If so, this would show up in the n-capture search and an excess of 
events after NC cuts would occur, thus finding the underestimated background. 

\clearpage
\section{Implementation}
In this section, we explore some details of how to prepare and run MiniBooNE with the addition
of scintillator.  

\subsection{Suggested plan for adding scintillator}
As of this writing, we plan to add 300~kg of PPO to the $1\times10^{6}$~liters of MiniBooNE mineral
oil (300~mg/l).  A preliminary price quote for PPO from the supplier to NOvA is \$250/kg  or \$75k for
scintillator.  The solubility of PPO will allow us to add the entire 300~kg to the MiniBooNE 
10~kl overflow and then introduce that into the main volume by recirculation.  However, it would be
prudent to do this addition in at least two steps by taking the concentration to about 50\% of the 
desired amount and
monitoring detector response with cosmic muons and muon-decay electrons.  We may do recirculation
without the addition of scintillator as a first step, 
as the MiniBooNE oil has not been recirculated since commissioning in 2002.

\subsection{Detector changes}
Our base plan for running with scintillator is only to add scintillator with no other changes.  New
readout electronics could be considered, but are not required for the physics goals set here.  The 
current rate of PMT and failures extrapolated for a 3-year run is not a problem.  We estimate that
the rate of electronics failures over that time period will be covered with our current supply 
of spares.  There will likely be some changes to the computing infrastructure to keep up with
hardware failures and security concerns, but an ``as-needed'' approach is our current plan. 

\subsection{Run plan}
As shown in Sec.~\ref{sec:analysis}, the neutron fraction measurement for oscillation candidates is
statistics limited and $6.5\times10^{20}$ POT is required for our current desired accuracy.  When the MicroBooNE
experiment is running, our assumption is that $2\times10^{20}$~POT/year will be delivered to the Booster Neutrino 
Beamline (BNB).  This sets a 3-year duration for the proposed scintillator phase of MiniBooNE. 
Preparation for running only requires the addition of scintillator (along with modest detector
maintenance), which we estimate will require about 3 months with no beam requirement.  If approvals and
funding can be obtained reasonably quickly, we estimate that we could be ready to add scintillator
by the end of 2013.

\subsection{Future Work}
There are several tasks and milestones to be done before the MiniBooNE collaboration is completely
prepared to go forward with adding scintillator.  Probably the most important task is to
show that the signal to background ratio in the oscillation search is not damaged by the 
addition of scintillator.  The results on electron reconstruction shown here are favorable evidence
that this can be accomplished and the LSND was able to do such a reconstruction, albeit at
lower energies.  However, the definitive test is to also show that particle identification
of $e$, $\mu$, $\pi^0$  is sufficient to repeat the $\nue$ appearance search with adequate background
rejection. Work on this in ongoing.

In addition, there are other tasks related to adding scintillator such as:
\begin{itemize}
\item adjustment of the MC simulation to account for the new light production
mechanism,  
\item additional light output tests, perhaps in 200~MeV proton beam,
\item light attenuation length tests,    
\item material compatibility tests,
\item scintillator procurement plan and price estimates, and 
\item detailed scintillator mixing and oil recirculation plan.

\end{itemize}
Work on these tasks is currently underway.

\clearpage
\section{Conclusions and request}
The addition of $300$~mg/l of scintillator to the existing MiniBooNE 
mineral oil will allow for the detection and reconstruction of 2.2~MeV $\gamma$  
from neutron-capture.  CC oscillation signal events should have an associated
neutron in less that 10\% of events in contrast to NC background events in which
$\approx 50\%$ have neutrons.   The neutron-capture 
rate for both of these event types can be separately measured in MiniBooNE, 
thus eliminating dependence on neutron production model calculations.  
Therefore, a measurement of neutron-capture in
oscillation events measures the NC backgrounds.  

A measurement of the neutron-fraction in a new appearance oscillation 
search with MiniBooNE will increase the significance of the oscillation
excess, if it maintains in the new data set, to $\approx 5\sigma$.  In practice,
the $\nue$ appearance search will be conducted again after the introduction
of scintillator.  With  $6.5\times10^{20}$ POT, the results of this search (before 
neutron capture cuts) should have similar sensitivity as existing search
but with different systematic errors.  Combining this with the neutron
capture analysis will raise the sensitivity to the $5\sigma$ level, perhaps
better, depending on final systematics.

This new phase of MiniBooNE would enable additional important 
studies such as the spin structure of nucleon ($\Delta s$) via NC elastic
scattering, a low-energy measurement of the neutrino flux via the 
$\numu~^{12}C \rightarrow \mu^{-}~^{12}N_\textrm{g.s.}$ reaction, and a test of the
quasielastic assumption in neutrino energy reconstruction.  
This effort will provide traning for  Ph.D.~students and postdocs and 
will yield important, highly-cited results over the next 5 years for 
a modest cost.

\begin{quotation}
\textit{This program of measurements requires approximately $6.5\times10^{20}$ 
protons on target delivered to MiniBooNE  and can begin in early 2014. 
We are requesting support of this concept to enable the collaboration 
to plan the experiment and analysis in more detail with the goal of submitting a full 
proposal for the experiment in mid-2013.}
\end{quotation}

\end{document}